\documentclass[a4paper,11pt]{article}
\usepackage{pos}

\newcommand\PhiBJ{\Phi_{\scriptscriptstyle \rm HJ}}

\newcommand{\noun}[1]{{\scshape #1}}

\newcommand{\POWHEG}{\noun{Powheg}}
\newcommand{\mcnlo}{\noun{MC@NLO}}

\newcommand{\minnlo}{{\noun{MiNNLO$_{\rm PS}$}}}

 \allowdisplaybreaks

 \newcommand{\dd}{\mathop{}\!\mathrm{d}}

\def\to{\rightarrow}

\newcommand{\eqn}[1]{Eq.\,(\ref{#1})}

\def\citere#1{\mbox{Ref.~\cite{#1}}}

\newcommand{\bbH}{\ensuremath{b\bar{b}H}}

\title{NNLO+PS predictions for Higgs production through bottom-quark fusion}
\author[a]{Christian Biello}
\author*[a,b]{ Aparna Sankar}
\author[a]{ Marius Wiesemann}
\author[a,b]{  Giulia Zanderighi}

\affiliation[a]{Max-Planck-Institut f\"ur Physik, \\Boltzmannstraße 8, 85748 Garching, Germany}

\affiliation[b]{Physik-Department, Technische Universit\"at M\"unchen, \\ James-Franck-Strasse 1, 85748 Garching, Germany}

\emailAdd{biello@mpp.mpg.de}
\emailAdd{aparna@mpp.mpg.de}
\emailAdd{marius.wiesemann@mpp.mpg.de}
\emailAdd{zanderi@mpp.mpg.de}

\abstract{We present next-to-next-to-leading-order (NNLO) QCD corrections for Higgs production through bottom-quark annihilation (\bbH{}) matched to parton showers (NNLO+PS) using the \minnlo{} technique. The \minnlo{} method is adapted for the extra scale dependence due to the Yukawa coupling renormalized in the $\overline{\rm MS}$ scheme. The computation has been carried out in the five flavour scheme (5FS) neglecting the bottom mass. Results are compared against fixed-order predictions at NNLO and resummed predictions at next-to-next-to-leading-logarithmic (NNLL) accuracy. We also present preliminary results within the four-flavour scheme (4FS) setup, reaching a new level of precision in the massive scheme.}

\FullConference{31st International Workshop on Deep Inelastic Scattering (DIS2024)\\
 8–12 April 2024\\
Grenoble, France\\}

\begin{document}
\maketitle

%----------INTRO-------------------------------------
\section{Introduction}
The Higgs boson is a key part of the Standard Model (SM) of particle physics. Since its discovery a decade ago, measuring its properties has been a primary focus at the Large Hadron Collider (LHC). These measurements confirm the SM and explore new physics. While current measurements of Higgs couplings align with SM predictions, more precise measurements could reveal small deviations. 
Accurate simulations of all Higgs production and decay modes at the LHC are essential for detecting deviations from SM predictions. The \bbH~process, though having a lower rate, is significant for precision measurements and beyond-the-SM (BSM) scenarios. It is challenging to detect directly due to large backgrounds and reduced rates when tagging bottom quarks but is crucial as an irreducible background in Higgs-boson pair production searches, especially at the High-Luminosity LHC (HL-LHC).

Calculations of the \bbH~process use either a five-flavour scheme (5FS) with massless bottom quarks or a four-flavour scheme (4FS) with massive bottom quarks. Significant advancements have been made in 5FS calculations, particularly with third-order QCD cross sections \cite{Duhr:2019kwi}. Progress in 4FS calculations has been slower due to their complexity, though NLO+PS in QCD combined with electroweak (EW) corrections remain state-of-the-art \cite{Wiesemann:2014ioa,Pagani:2020rsg}. Matching of NNLO QCD calculations with parton showers (NNLO+PS) presents a significant challenge in collider theory. For the \bbH~process, the first matching was done at NLO in 4FS using the \mcnlo{} and \POWHEG{} methods. Here, we present the first fully-differential NNLO QCD calculation matched to a parton shower using the \minnlo{} method  \cite{Monni:2019whf,Monni:2020nks} within the \POWHEG{} framework \cite{Biello:2024vdh}.
%----------------------------------------------------------
\section{Theoretical framework}
The \minnlo{} cross section for $b\bar{b}\to H$ production can be expressed through the standard \POWHEG{} formula 
for $HJ$ production with a modified content of the \POWHEG{} $\bar{B}$ function \cite{Nason:2004rx}:
\begin{align}
   \dd \sigma_{\text{\scalefont{0.77}H}}^{\text{\minnlo{}}} &= \dd \PhiBJ \bar{B}^{\text{\minnlo{}}} \times \Big\{ \Delta_{\text{pwg}}(\Lambda_{\text{pwg}}) 
   + \dd \Phi_{\text{rad}} \Delta_{\text{pwg}}(p_{\text{T,rad}}) \frac{R_{\text{\scalefont{0.77}HJ}}}{B_{\text{\scalefont{0.77}HJ}}}  \Big\} \,, \label{bpwg}
\end{align}
where $\Phi_{\rm HJ}$ is the $HJ$ phase space. $\Delta_{\text{pwg}}$ is the \POWHEG{} Sudakov form factor with a cutoff $\Lambda_{\text{pwg}}=0.89\text{ GeV}$. $\dd \Phi_{\text{rad}}$ and $p_{T,\text{rad}}$ denote the phase space measure and transverse momentum of the real radiation relative to $HJ$ production. \POWHEG{} matches the fixed-order $HJ$ calculation with a parton shower by generating the first additional radiation using the ratio of the tree-level matrix elements for $HJJ$ ($R_{\text{\scalefont{0.77}HJ}}$) and $HJ$ ($B_{\text{\scalefont{0.77}HJ}}$) productions. Subsequent radiations with smaller transverse momenta are generated by a Shower Monte Carlo.

The central ingredient of the \minnlo{} method is the modified \POWHEG{} $\bar B$ function which is denoted as $\bar{B}^{\text{\minnlo{}}}$ in \eqn{bpwg}. The derivation of this function stems from transverse momentum resummation of color singlet production. $\bar{B}^{\text{\minnlo{}}}$ function is given by \cite{Monni:2019whf,Monni:2020nks}  
\begin{align}
  \bar{B}^{\text{\minnlo{}}}&=e^{-\tilde S(p_{T})}  \bigg\{ B \left( 1 + \tilde{S}^{(1)} \right) + V + \int \mathrm{d\Phi_{rad}} ~R + D^{(\ge 3)}(p_T)
  \times F^{\text{corr}} \bigg\}\,, \label{bminnlo}
\end{align}
where Born $B$, virtual $V$ and the real $R$ contributions are calculated with strong coupling at $p_T$, transverse momentum of the Higgs. Here $e^{-\tilde S(p_{T})}$ is the Sudakov form factor which regulates the divergence of the  $\bar{B}^{\text{\minnlo{}}}$ function for $p_T \rightarrow 0$ and $\tilde{S}^{(1)}$ is its first order expansion coefficient.  $D^{(\ge 3)}(p_T)$ term multiplied with the spreading function $F^{\text{corr}}$ contains the additional pieces which are necessary to attain the NNLO accuracy for the inclusive observables for the color singlet. 

%-----RESULTS----------------------------------------------
\section{Phenomenological results for the 5FS}

\begin{figure*}[t!]
\begin{center}
 \begin{tabular}{cc}
\includegraphics[width=.370\textwidth]{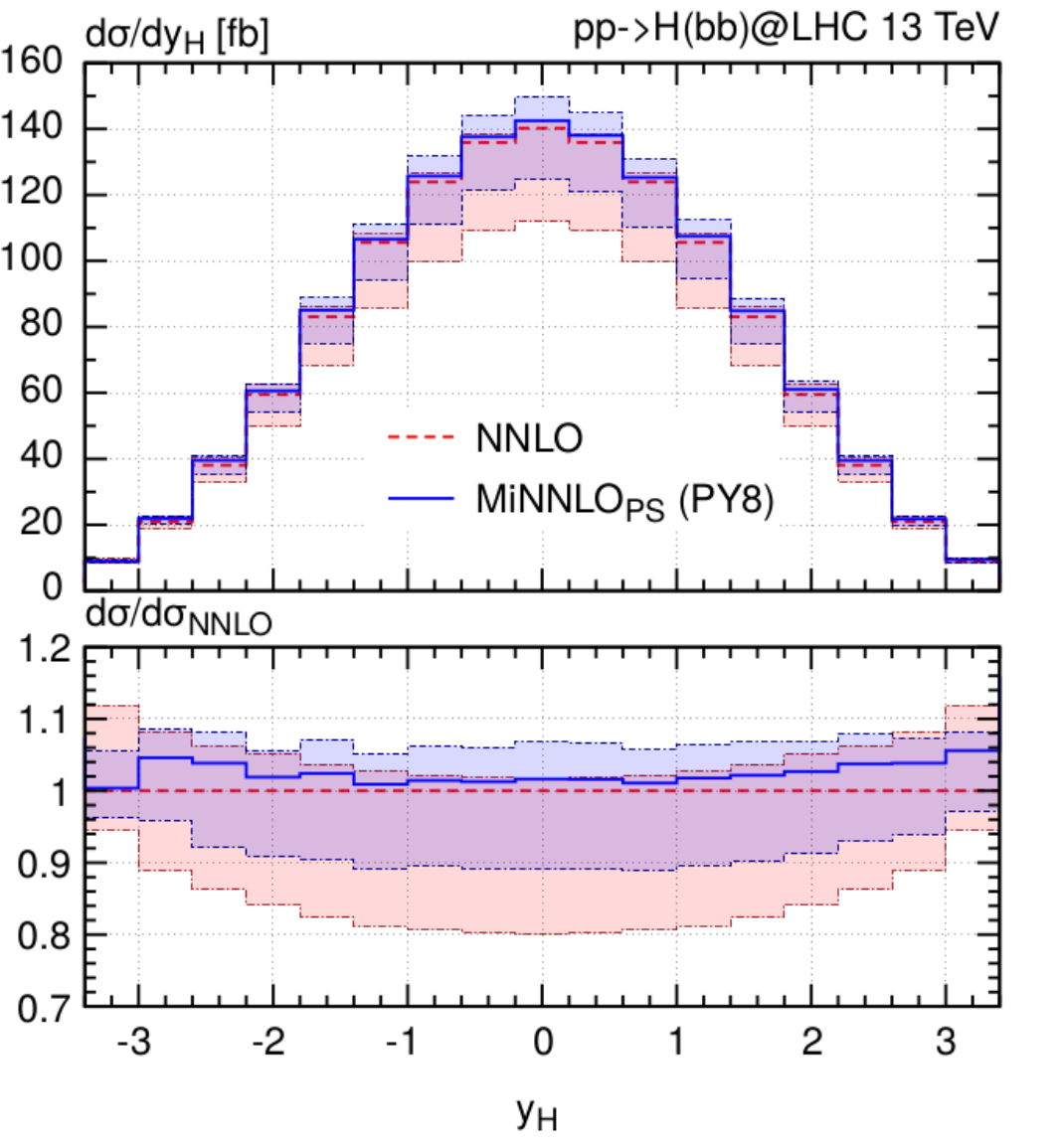}
 \hspace{-0.06cm}
 \includegraphics[width=.364\textwidth]{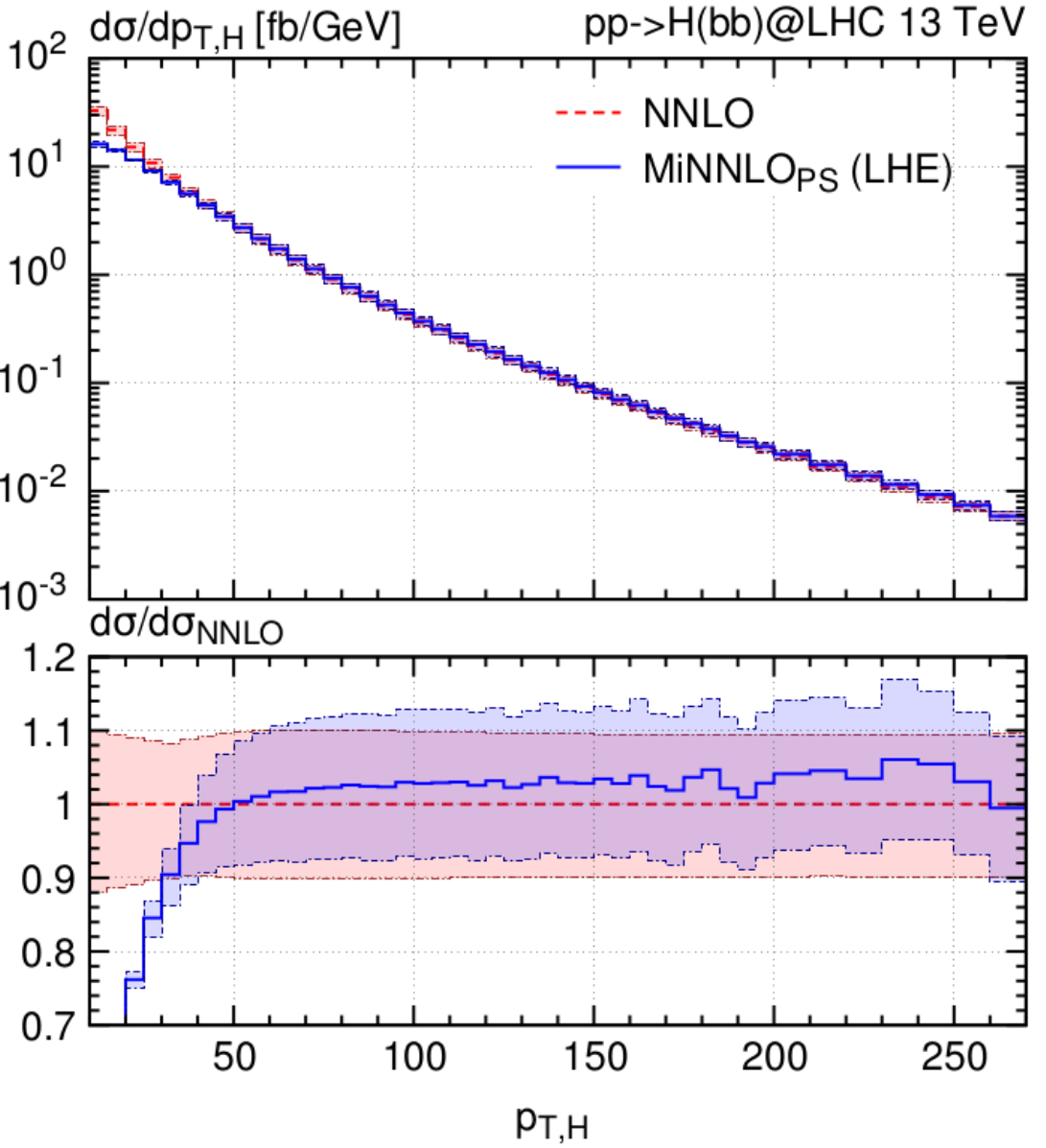}
% \includegraphics[width=.45\textwidth]{fi2.pdf}
% &
% \hspace{-0.8cm}
% \includegraphics[width=.45\textwidth]{defplots/figure2b.pdf}
 \end{tabular}
% \vspace*{1ex}
 \caption{The left plots show the comparison of \minnlo{} predictions (blue, solid) with the NNLO results of \citere{Mondini:2021nck}
(red, dashed) for the rapidity distribution of the Higgs boson. The right plot shows the analytic $p_{T,H}$ spectrum 
up to $\alpha_s^2$ from \citere{Harlander:2014hya} (red, dashed) and our \minnlo{} prediction (blue, solid curve).  \label{fig:1}}
\end{center}
\end{figure*}
%-------------------------------TEXT------------
We present predictions for Higgs boson production via bottom-quark annihilation at the LHC (13 TeV), with a Higgs boson mass of 125 GeV. Using the 5FS approach with massless bottom quarks and a non-zero Yukawa coupling in the $\overline{\rm MS}$ scheme, we employ the NNLO NNPDF40 parton distribution functions (PDFs) with $\alpha_s$($m_Z$) = 0.1189 and assess theoretical uncertainties via a standard 7-point scale variation. For the simulations with parton shower, we use Pythia8 (PY8) with the A14 tune, excluding effects like hadronization, multi-parton interactions (MPI), and QED radiation.

Employing the \minnlo{} method, we obtain an inclusive cross section of $0.509_{-5.3\%}^{+2.9\%}$\,pb which is consistent with the NNLO prediction of $0.518_{-7.5\%}^{+7.2\%}$\,pb from {\sc SusHi} \cite{Harlander:2012pb,Harlander:2003ai}. We further validate our \minnlo{} generator by comparing the rapidity and transverse momentum distributions of the Higgs against the corresponding fixed-order results \cite{Mondini:2021nck,Harlander:2014hya} in Fig.\ref{fig:1}. Using the NNLO set of CT14 PDFs, we find good agreement in the rapidity distribution, as shown in Fig.\ref{fig:1}. For the $p_{T,H}$ spectrum, NNLO and \minnlo{} results agree at higher $p_{T,H}$ values. However, for $p_{T,H} \rightarrow 0$, \minnlo{} remains finite whereas NNLO diverges.

\begin{figure*}[t!]
\begin{center}
 \begin{tabular}{cc}
\includegraphics[width=.354\textwidth]{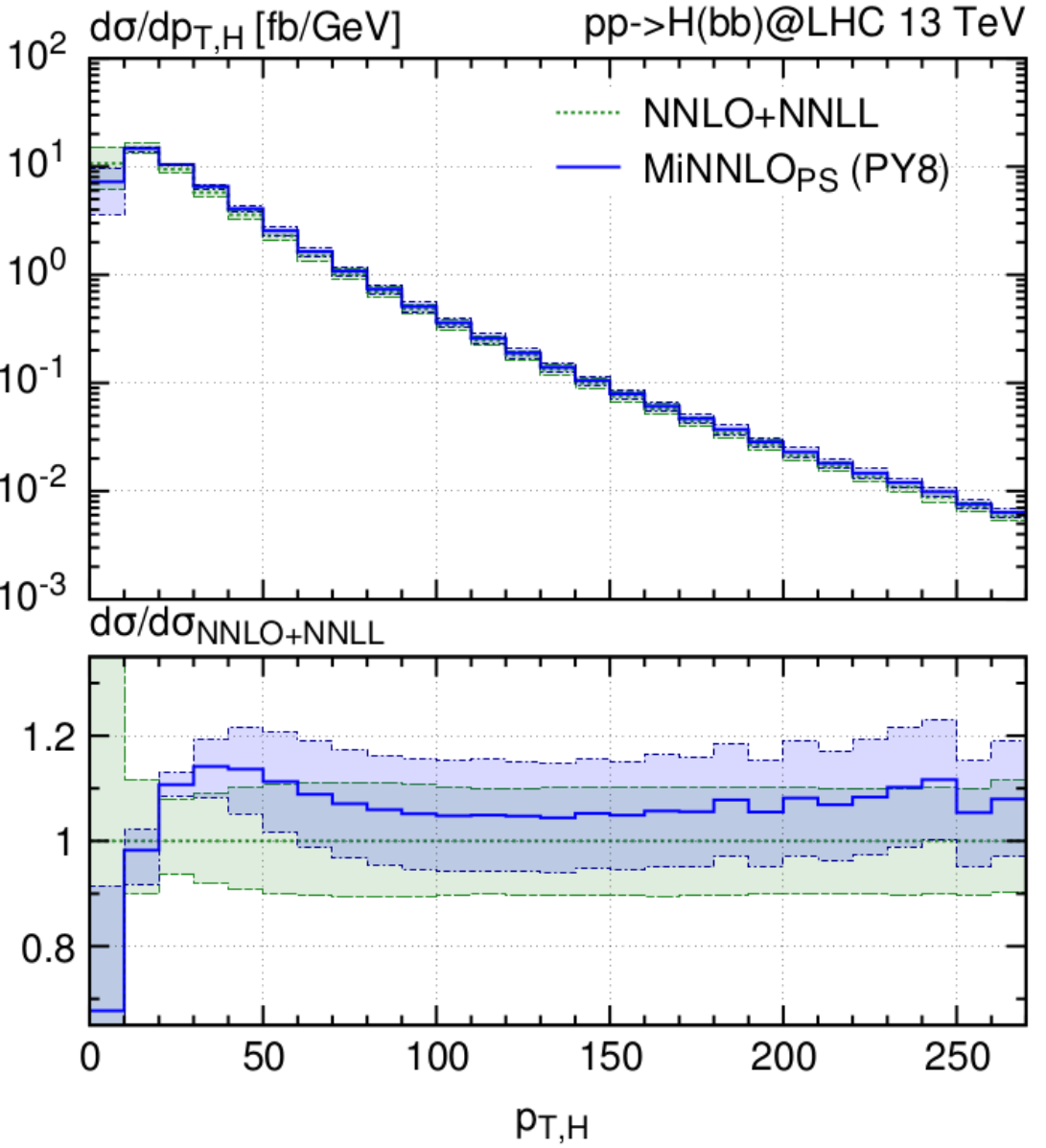}
 \hspace{-0.06cm}
 \includegraphics[width=.354\textwidth]{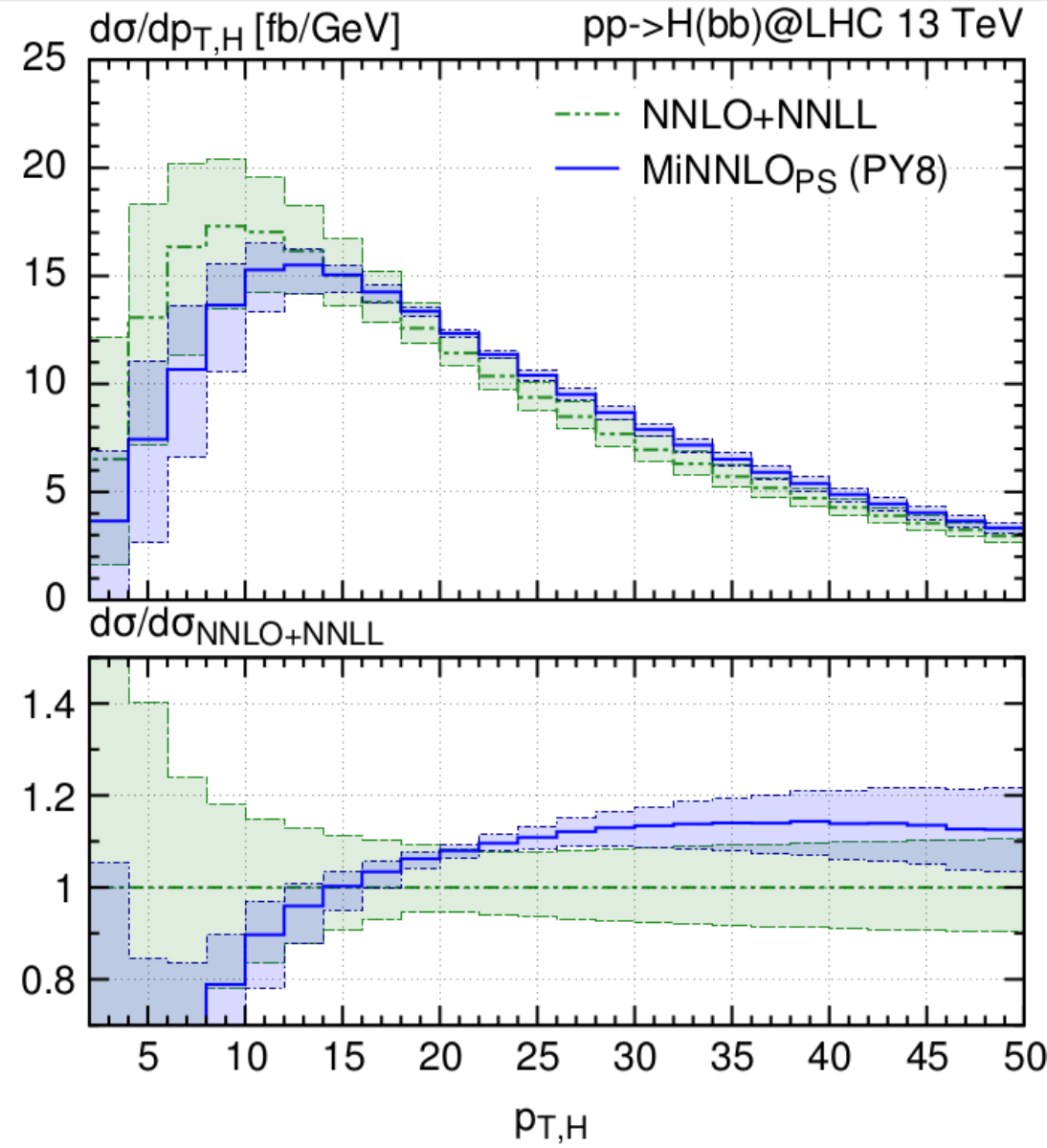}
% \includegraphics[width=.45\textwidth]{fi2.pdf}
% &
% \hspace{-0.8cm}
% \includegraphics[width=.45\textwidth]{defplots/figure2b.pdf}
 \end{tabular}
% \vspace*{1ex}
 \caption{Predictions for the $p_{T,H}$ spectrum from our \minnlo{} generator compared to the analytic resummation at NNLO+NNLL of \citere{Harlander:2014hya} (green, double-dash-dotted curve)\label{fig:pthresum}.}
\end{center}
\end{figure*}
Fig.\ref{fig:pthresum} compares \minnlo{} predictions for the $p_{T,H}$ distribution, including parton showering effects from PYTHIA8, against the analytic resummation at NNLO+NNLL \cite{Harlander:2014hya}. At large $p_{T,H}$, both results agree within the uncertainty band. However, at small $p_{T,H}$, \minnlo{} scale bands underestimate uncertainties compared to more accurate NNLO+NNLL, suggesting the need for further variations within the shower settings to study this effect.
%--------------------4FS--------------------------------
\section{Preliminary results for the 4FS}
The 4FS implementation of the \minnlo{} generator for \bbH~production presents greater complexity due to the higher final state multiplicity and the inclusion of massive colored final states. We provide preliminary results for the Higgs total cross section, employing the \minnlo{} method for heavy-quark pair production plus a color singlet \cite{Mazzitelli:2024ura}, accounting for the Yukawa scale dependence. While the two-loop amplitude for Higgs associated with massive bottom quarks is not known, we approximate it by applying a massification procedure \cite{Mitov:2006xs,Wang:2023qbf} to the massless amplitude \cite{Badger:2021ega}, effectively capturing leading mass effects.

By setting the Higgs mass for the Yukawa scale and using a quarter of the transverse invariant mass for the  strong couplings at the Born level, the NLO+PS generator predicts a cross-section of $0.381^{+20\%}_{-16\%}$ pb. This value is approximately 55\% smaller than the NLO cross-section in 5FS, indicating the importance of collinear logarithmic contributions. On the other hand, \minnlo{} predicts a cross-section of $0.464^{+14\%}_{-13\%}$ pb, which is much closer to NNLO 5FS values. This marks the first time that predictions from both schemes have shown agreement within the scale uncertainty.
%-------------CONCLUSION-----------------------------
\section{Conclusions}
We presented the NNLO QCD matching with parton shower for Higgs production via bottom fusion using the \minnlo{} technique. Our analysis covers predictions in the 5FS and discusses ongoing investigations in the 4FS \cite{our4FS}, showing promising preliminary findings.
%------------ACK-------------------
% \section{Acknowledgements}
% % We would like to thank Ciaran Williams for providing us with numbers for the 
% % NNLO rapidity distribution of the Higgs boson in $b\bar{b}\to H$ production.
% We have used the Max Planck Computing and Data Facility (MPCDF) in Garching to carry 
% out all simulations presented here.
%--------------------BIB-----------------------------------
\bibliographystyle{jhep}
\bibliography{bbH}

\end{document}